\documentclass[english,ngerman]{bvm} 

\bvmdef\articlenumber{3208}

\bvmdef\type{P}

\date{}

\title{Multiscale Softmax Cross Entropy for Fovea Localization on Color Fundus Photography}

\titlerunning{Multiscale Softmax Cross Entropy}
\author{Yuli~Wu$^1${\href{https://orcid.org/0000-0002-6216-4911}{\includegraphics[height=9pt]{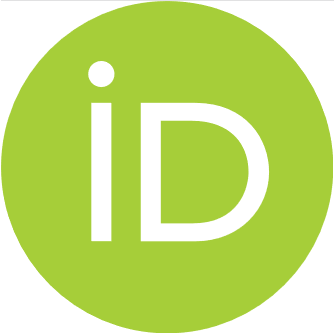}}}, Peter~Walter$^2$, Dorit~Merhof$^1$}

\authorrunning{Wu, Walter \& Merhof}

\institute{%
$^1$Institute of Imaging and Computer Vision, RWTH Aachen, Germany \\
$^2$Department of Ophthalmology, RWTH Aachen, Germany
}

\email{yuli.wu@lfb.rwth-aachen.de}
\makeatletter
\let\temp@caption\@caption  
\usepackage[colorlinks=true]{hyperref}
\let\@caption\temp@caption 

\begin{document}

%
\selectlanguage{english}

\maketitle
\begin{abstract}
Fovea localization is one of the most popular tasks in ophthalmic medical image analysis, where the coordinates of the center point of the \textit{macula lutea}, i.e. \textit{fovea centralis}, should be calculated based on color fundus images. In this work, we treat the localization problem as a classification task, where the coordinates of the x- and y-axis are considered as the target classes. Moreover, the combination of the softmax activation function and the cross entropy loss function is modified to its multiscale variation to encourage the predicted coordinates to be located closely to the ground-truths. Based on color fundus photography images, we empirically show that the proposed multiscale softmax cross entropy yields better performance than the vanilla version and than the mean squared error loss with sigmoid activation, which provides a novel approach for coordinate regression.
\end{abstract}

\section{Introduction}
Outputting coordinates is common in computer vision tasks, such as in bounding-box regression for object detection and in keypoint localization for facial recognition. Regression losses are typically selected to calculate the error between the ground-truth and the prediction, \textit{e.g.} Mean Squared Error (MSE) loss and Mean Absolute Error (MAE) loss, which measure L2 and L1 distances between the ground-truth and the prediction, respectively. In contrast, probabilistic losses are usually used in classification tasks, which includes Cross Entropy (CE) loss as one of the most popular choices (categorical cross entropy in the case of multi-class). One significant difference between these two categories is that MSE or MAE punishes incorrect predictions less, which are however close to the ground-truth, while categorical CE combined with softmax activation function treats all incorrect predictions equally to the maximum.

An accurate localization of the fovea, an important anatomical landmark in the retina, can be beneficial to the computer aided diagnosis of retinal diseases. Huang et al.~\cite{huang2020efficient} take advantage of the geometrical relationship between optic disc and fovea to achieve a more accurate localization and Xie et al.~\cite{XieEndtoEnd2021} utilize a three-stage network with coarse-fine fusion. MSE loss is used in both approaches. Kopaczka et al.~\cite{kopaczka2020robust} combine soft-argmax loss~\cite{honari2018improving} and L1 distance loss to localize and track rodents, which shows the feasibility of probabilistic loss functions in classification tasks.

In this work, we consider the localization task as two classification tasks, referring to the x- and y-axis, by using a combination of the softmax activation function and the cross entropy loss function. Trying to bridge the functional gap between the regression and probabilistic losses, we propose Multiscale Softmax Cross Entropy (MSCE), which takes the last feature map learned from the backbone convolutional neural network and combines multiple downsampled feature maps with independent softmax cross entropy.

\begin{figure}[h]
	\setlength{\figbreite}{0.98\textwidth}
	\centering
	\includegraphics[width=\figbreite]{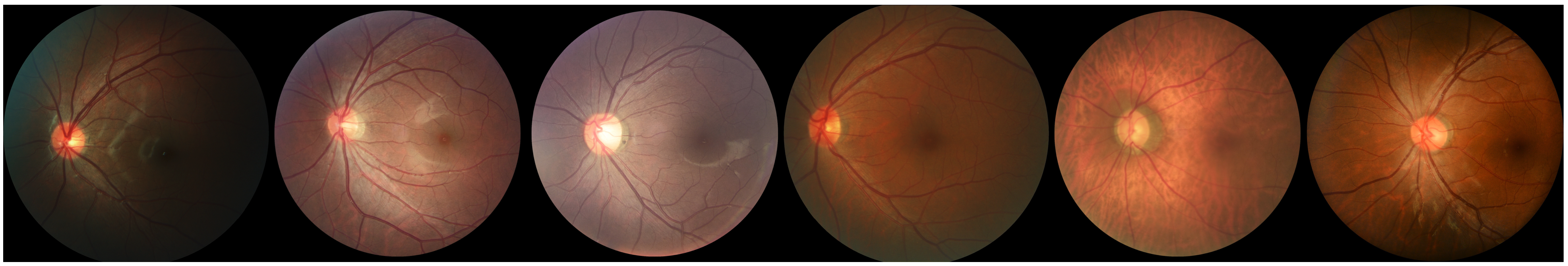}
	\label{retina}
	\caption{Examples of color fundus images from REFUGE2~\cite{OrlandoREFUGE2020}.}
\end{figure}

\section{Materials and Methods}

\subsection{Dataset}
The dataset of color fundus images, REFUGE2~\cite{OrlandoREFUGE2020}, contains 1200 images with and 400 images without ground-truth annotations for training and testing, respectively. The metrics used to evaluate the predicted localization coordinates is taken from the latest Gamma Challenge\footnote{\url{https://gamma.grand-challenge.org}}, namely the Reciprocal of the Average Euclidean Distance (R-AED) value, which is defined as \(\text{R-AED} = \frac{1}{d(\mathbf{p},\mathbf{q})+0.1}\), where the Euclidean distance is used between the normalized coordinates of ground-truth \(\mathbf{p}\) and prediction \(\mathbf{q}\) as  \(d(\mathbf{p},\mathbf{q}) = ||\mathbf{p}-\mathbf{q}||_2\).

\subsection{Network Architecture}
We adopt the neural network architecture from cellpose~\cite{StringerCellpose2021}, which is a modified U-Net~\cite{ronneberger2015u} with residual connections inside each convolutional block and a style vector fused to the upsampling pathway. The original image is first resized and fed into the cellpose network, which outputs the feature map of identical size. The learned feature map is pooled multiple times to generate the multiscale branches, each of which is first reduced per axis (via \textit{e.g.} \texttt{sum} or \texttt{mean}). The multiscale loss is then calculated with independent softmax cross entropy for each branch. Finally, the final loss is aggregated with weighted sum, which we denote as Multiscale Softmax Cross Entropy (MSCE). The detailed introduction of MSCE is presented in Section~\ref{secloss}. The implementation of hyperparameters during the experiments can be found in Section~\ref{setup}.

\begin{figure}[t]
	\setlength{\figbreite}{\textwidth}
	\centering
	\includegraphics[width=\figbreite]{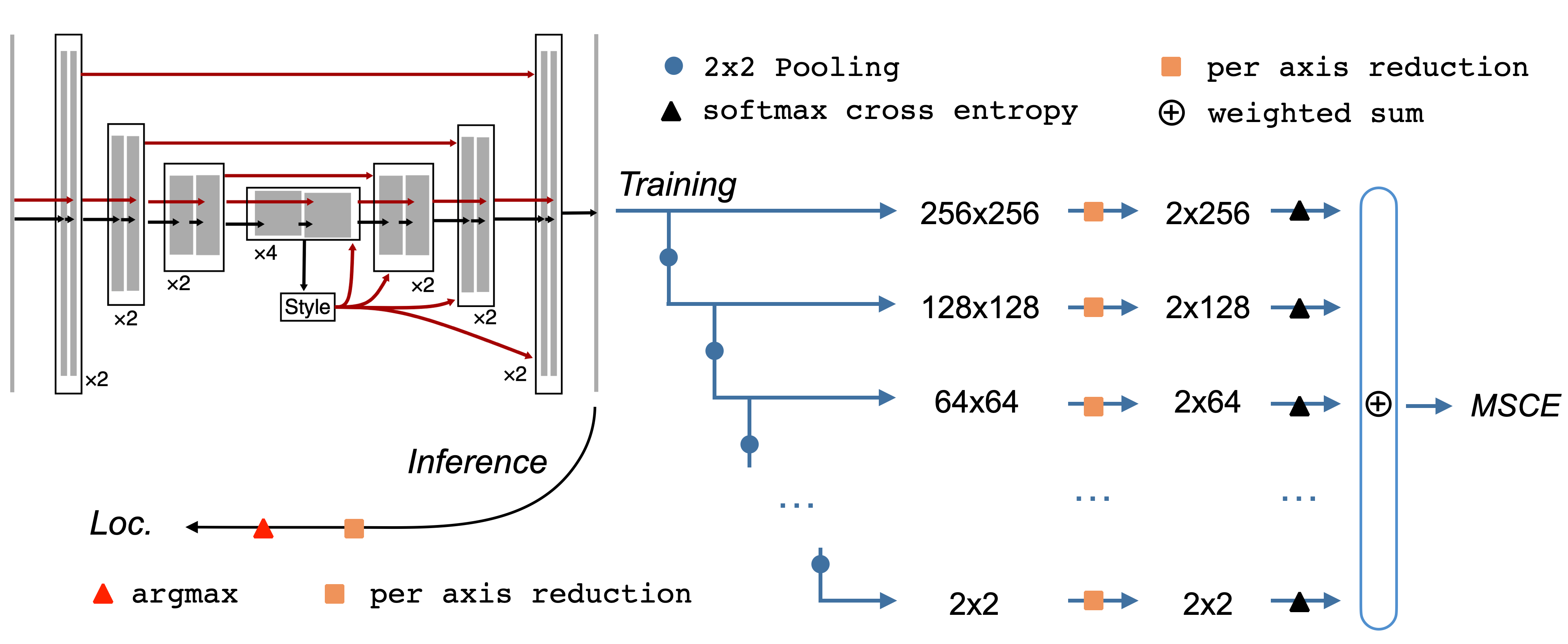}
    \caption{Network architecture for training and inference. The network backbone is adopted from the cellpose network~\cite{StringerCellpose2021} and the corresponding figure is adapted from it.}
	\label{network}
\end{figure}

\subsection{Loss}\label{secloss}
We present Multiscale Softmax Cross Entropy (MSCE), which takes two logit vectors of different sizes and calculates a weighted summation of softmax cross entropy from them.
\begin{equation}
    SCE = - \sum_{i=1}^{C} t_i \log( \frac{e^{s_i}}{\textstyle{\sum_{j=1}^{C}{e^{s_j}}}})
    \label{sce}
\end{equation}

\begin{equation}
    MSCE = \sum_{m=1}^{M} \lambda_m \cdot \big( - \sum_{i=1}^{C_m} t_i \log( \frac{e^{s_i}}{\textstyle{\sum_{j=1}^{C_m}{e^{s_j}}}}) \big)
     \label{msce}
\end{equation}

Based on the original Softmax Cross Entropy (SCE) in Equation~\ref{sce}, the multiscale version can be defined as shown in Equation~\ref{msce}. In both equations, \(s\) denotes the predicted logit and \(t_i\) indicates whether the \(i\)-th of total \(C\) class labels is the correct classification. In Equation~\ref{msce}, \(M\) denotes the number of multiscales (or the number of the branches in Fig.~\ref{network}) and \(\lambda_m\) denotes the weights for the SCE term of each scale. In this work, we set all \(\lambda_m = 1\).

\begin{figure}[t]
	\setlength{\figbreite}{0.45\textwidth}
	\centering
	\subfigure[MSE]{\includegraphics[width=\figbreite]{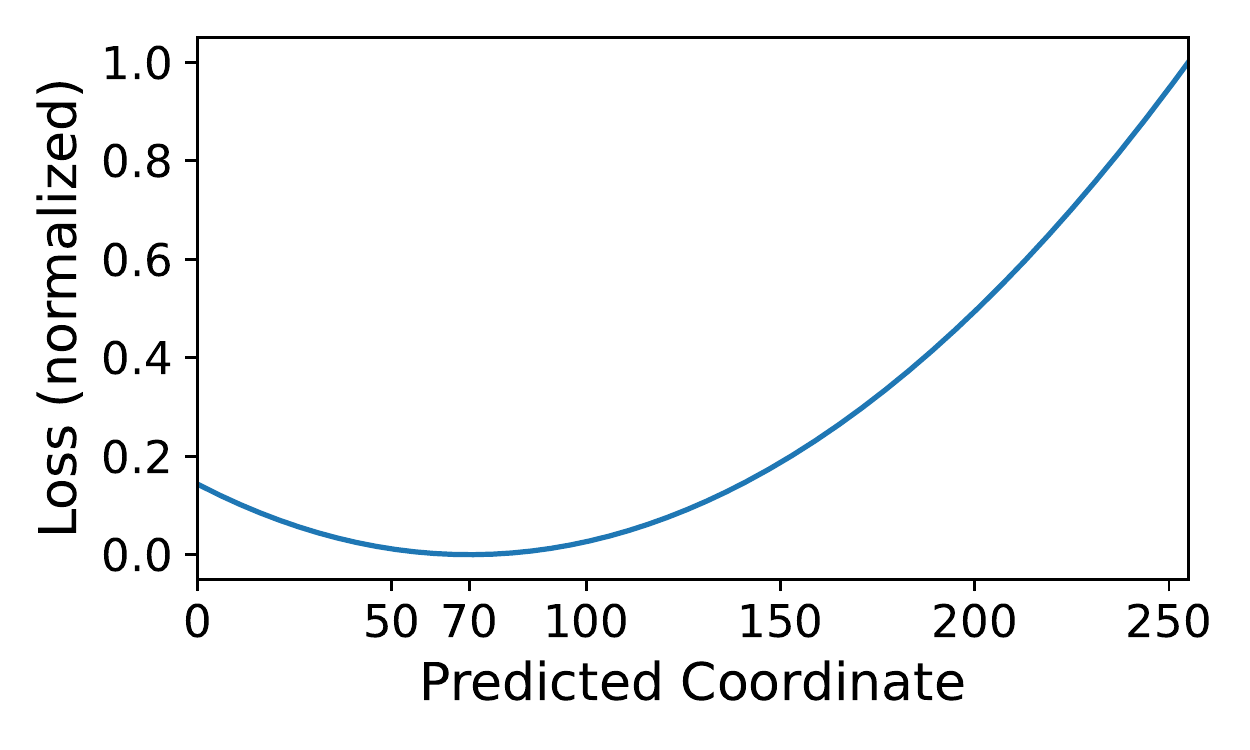}}
	\subfigure[SCE]{\includegraphics[width=\figbreite]{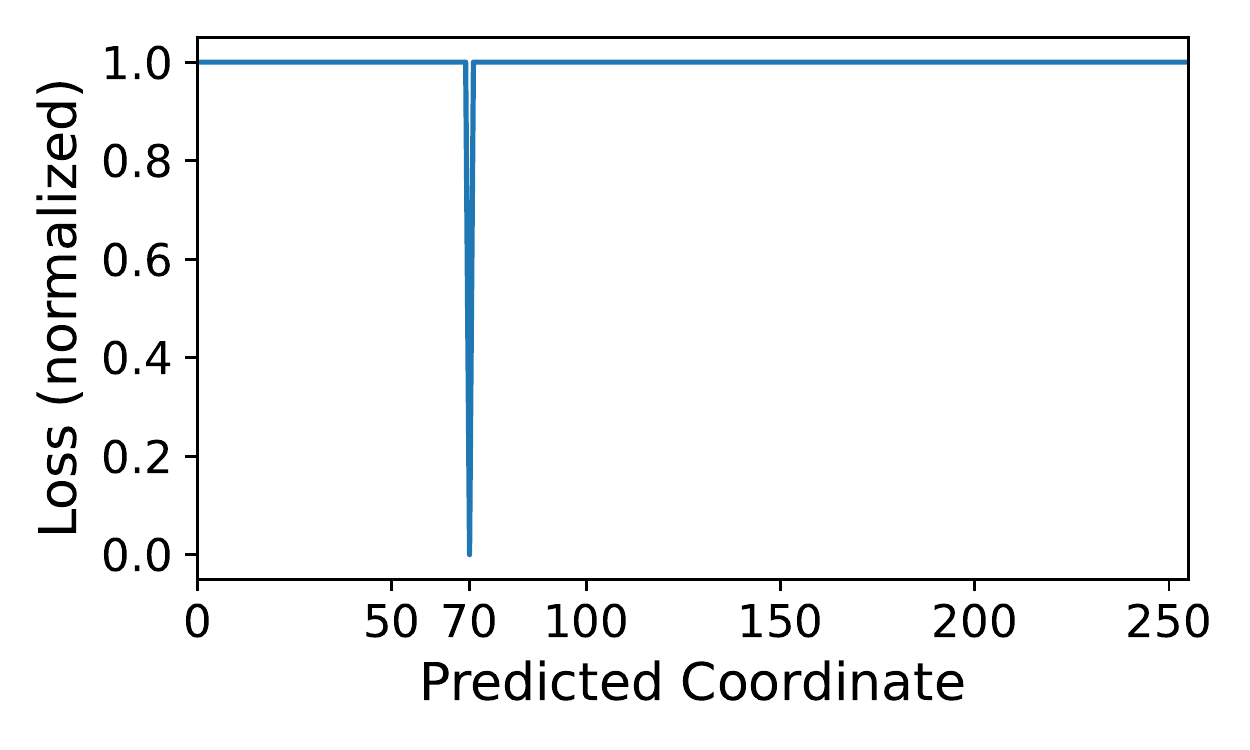}}\\\bigskip
	\subfigure[MSCE (\(M=4\))]{\includegraphics[width=\figbreite]{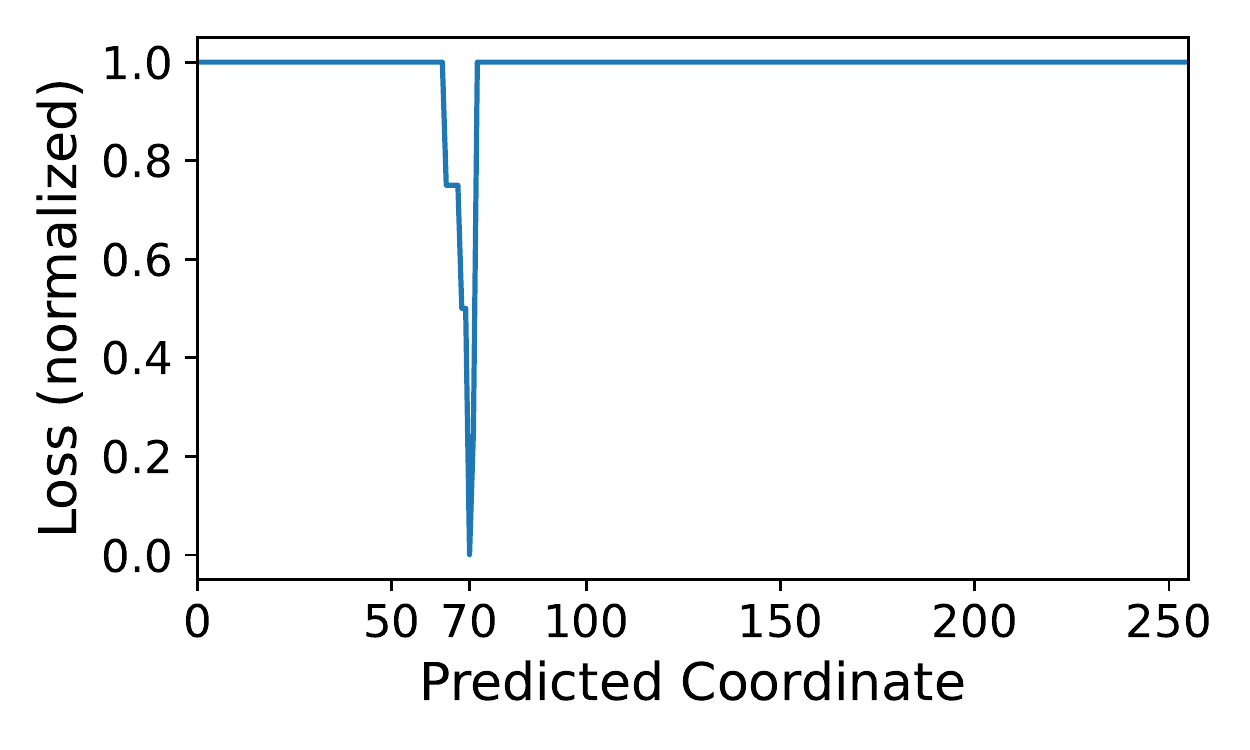}}	
	\subfigure[MSCE (\(M=8\))]{\includegraphics[width=\figbreite]{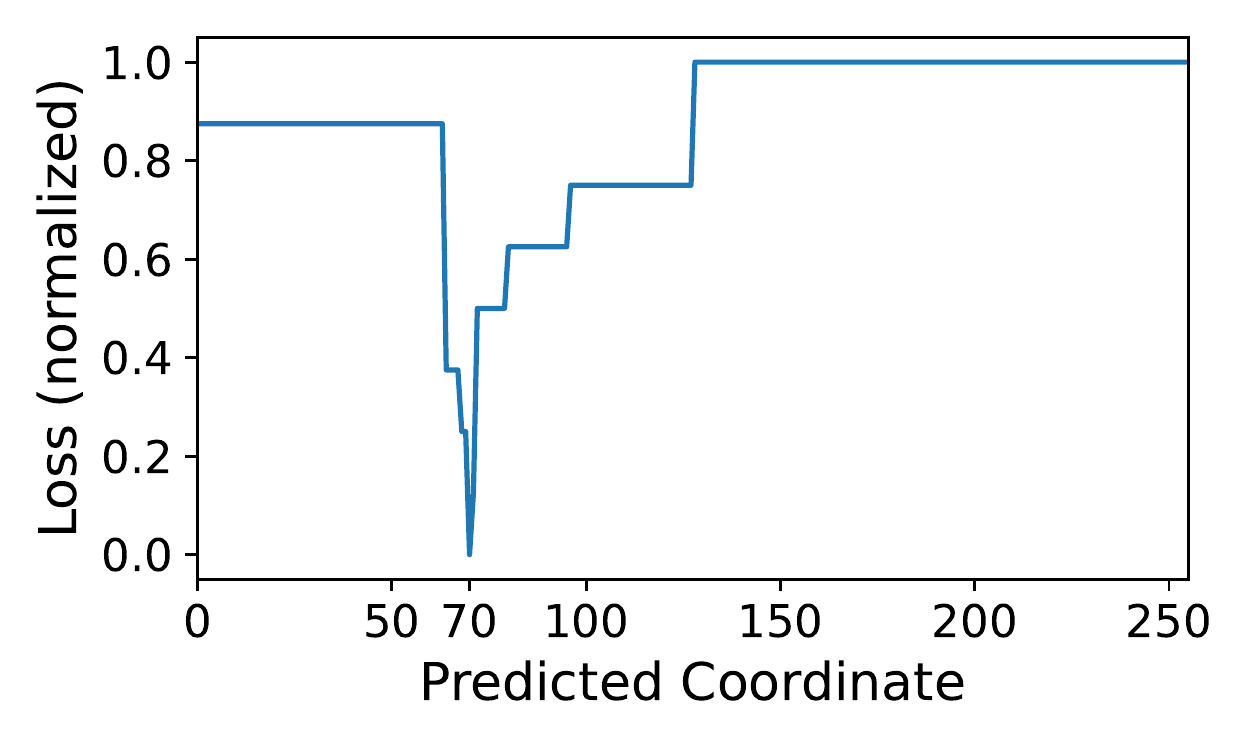}}
	\caption{Toy experiments in 1D coordinate: loss values of different loss functions. Predicted coordinates are presented in x-axis and the normalized loss values in y-axis, with the assumption that the 70th coordinate of total 256 is the ground-truth. MSE: Mean Squared Error; SCE: Softmax Cross Entropy; MSCE: Multiscale Softmax Cross Entropy. \(M\) denotes the number of the multiscales (see Equation~\ref{msce}). }
	\label{loss}
\end{figure}

In Fig.~\ref{loss}, different loss functions are compared with a toy example, where we assume the 70th class, \textit{i.e.} coordinate, of a 256 dimensional vector is the ground-truth and the normalized loss values are calculated for each possible prediction. Fig.~\ref{loss}(a) illustrates MSE, an example from the category of regression loss, which progressively attracts the wrong predictions to the ground-truth. In the case of SCE (Fig.~\ref{loss}(b)), however, the incorrect coordinates have been opposed expressly and unanimously, no matter where they are located rather than the ground-truth. The proposed MSCE is expected to neutralize the characteristics of MSE and SCE, which not only distinguishes the predictions in a stepwise regressive manner but also drastically encourages the prediction to converge towards the single actual ground-truth without decreasing the reward ratio. The desired feature can be better approached, if the number of the multiscales \(M\) is set to the maximum (\(M=8\) in case of 256 classes) comparing Fig.~\ref{loss}(c) and Fig.~\ref{loss}(d).

\subsection{Hyperparameters}\label{setup}
The \textit{style} mechanism from the cellpose network~\cite{StringerCellpose2021} has been preserved, as it is assumed to play a role when combining the disease grading task and the fovea localization task in future work. The images are first resized to 256 by 256 and no augmentation techniques have been then applied. We use \texttt{MaxPooling} when downsampling the feature map to generate multiscales ones (blue dots in Fig.~\ref{network}) and \texttt{sum} as the reduction operator to obtain the per axis logit vectors (orange squares in Fig.~\ref{network}). It has been empirically shown that \texttt{MaxPooling} and \texttt{sum} yield better results than \texttt{AveragePooling} and \texttt{mean} reduction.

The training process is optimized by stochastic gradient descent that uses an exponential decay schedule with an initial learning rate of 0.01, decay steps of 400 and a decay rate of 0.9. The maximal number of epochs is set to 1000 and the \texttt{EarlyStopping} mechanism has been applied with a patience of 100 epochs in terms of overall loss values.

\section{Results}
Experimental results are shown in Table~\ref{tab} \textit{w.r.t.} the reciprocal of the average Euclidean distance (R-AED). It is found that \texttt{MaxPooling} with \texttt{sum} reduction plays a significant role to boost the performance with SCE and MSCE, in which case MSCE outperforms MSE loss. Generally, the modified MSCE yields better results than the vanilla SCE, which empirically demonstrates the feasibility of probabilistic losses in the regression tasks. 

\begin{figure}[t]
	\setlength{\figbreite}{0.19\textwidth}
	\centering
	\subfigure[MSE]{\includegraphics[width=\figbreite]{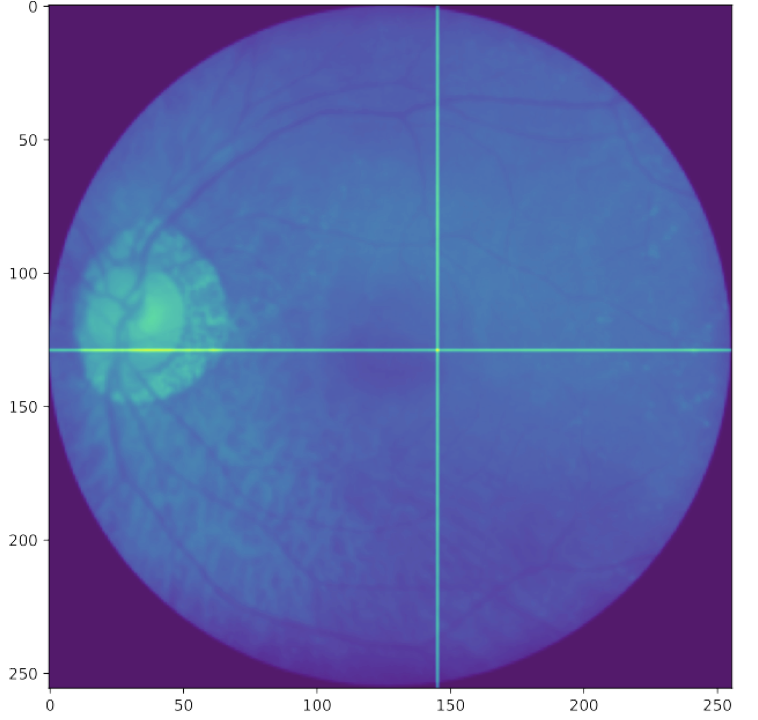}}
	\subfigure[SCE]{\includegraphics[width=\figbreite]{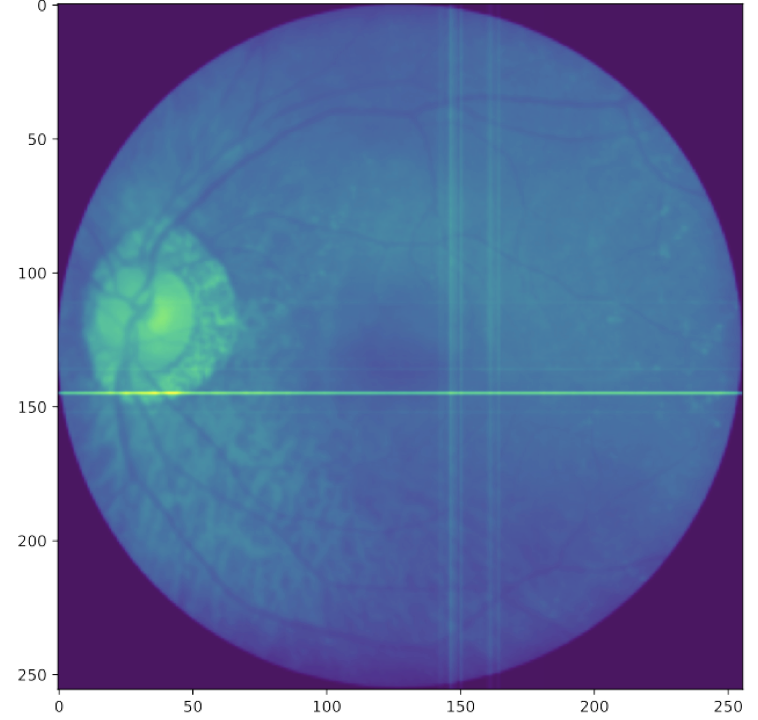}}
	\subfigure[MSCE]{\includegraphics[width=\figbreite]{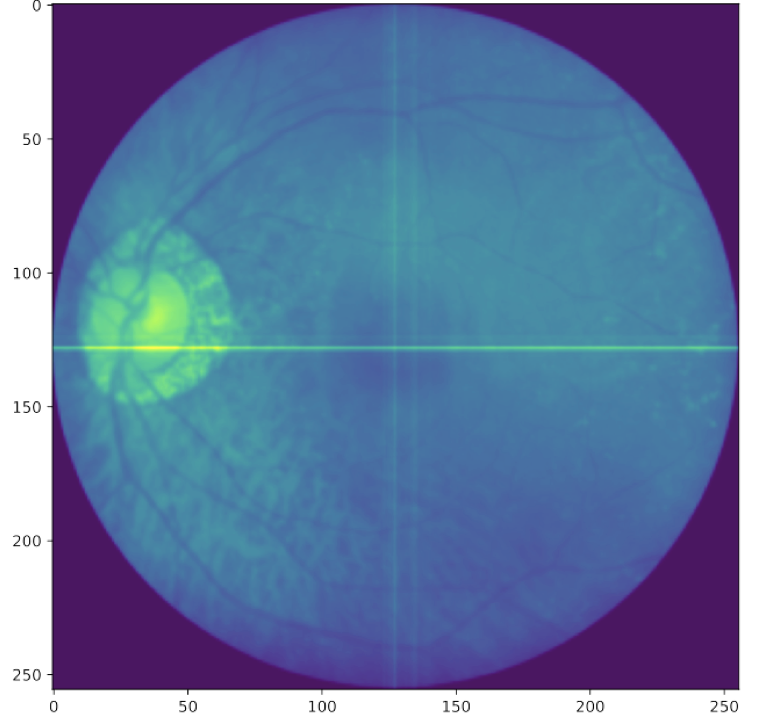}}
	\subfigure[Results]{\includegraphics[width=\figbreite]{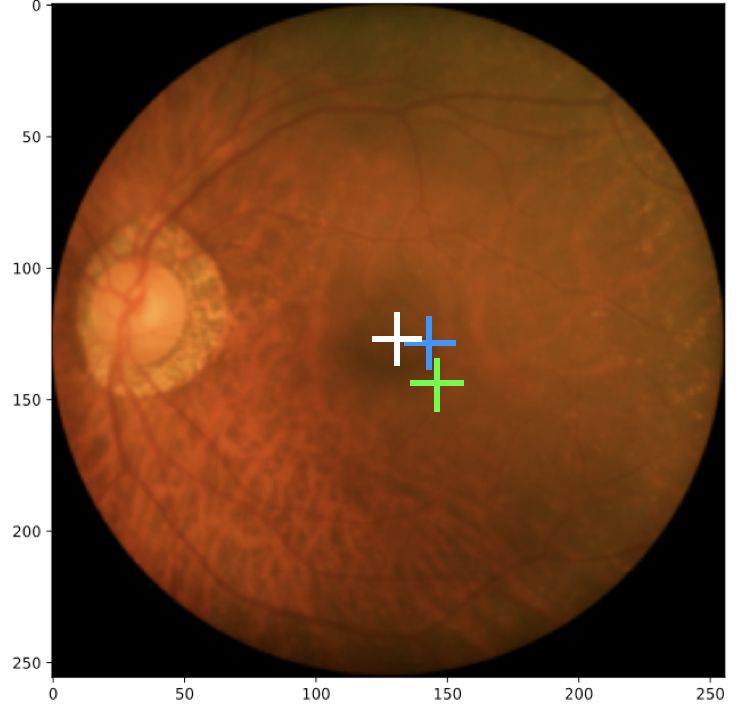}}
	\subfigure[MSCE failed]{\includegraphics[width=\figbreite]{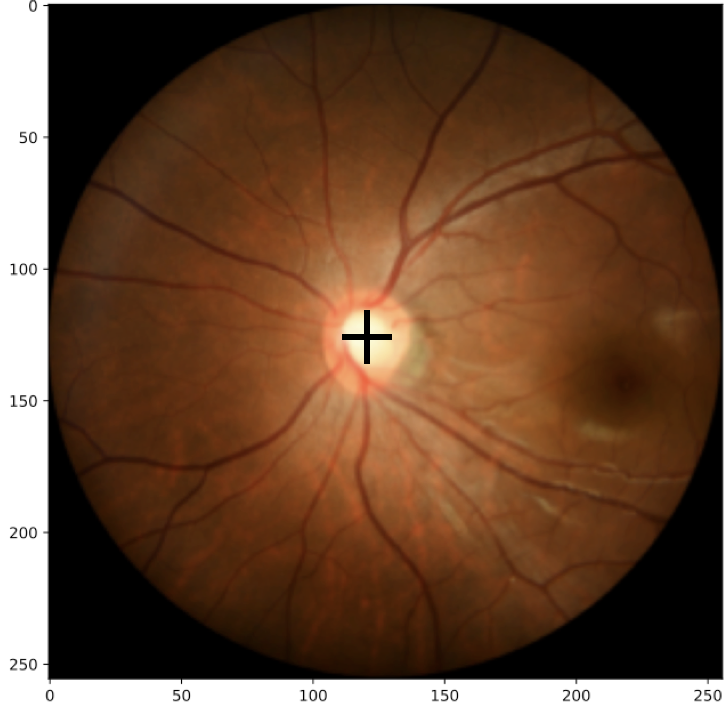}}
	\caption{Examples of predicted fovea locations with different losses. (a-c) illustrate the final coordinate vectors on the original images. (d) compares the predicted locations with crosses (blue, green, white denote MSE, SCE, MSCE, respectively). (e) shows a failed prediction, if the optic disc instead of the fovea is located in the center.}
	\label{results}
\end{figure}

Predicted fovea locations are illustrated in Fig.~\ref{results}, where the final coordinate vectors are illustrated on the original fundus images with mean squared error (MSE), vanilla softmax cross entropy (SCE) and multiscale softmax cross entropy (MSCE) in (a-c), respectively. From Fig.~\ref{results}(d), it can be noted that MSE (blue) and SCE (green) result in a larger offset than MSCE (white). A typical failed prediction happens if the fovea is located far away from the central region and blends into the dark marginal area, as shown in Fig.~\ref{results}(e). 

\begin{table}[t]
\centering
\caption{Results from ablation experiments with different loss functions and network settings \textit{w.r.t.} the reciprocal of the average Euclidean distance (R-AED). \texttt{Ave/mean} denotes \texttt{AveragePooling} with \texttt{mean} reduction and \texttt{Max/sum} denotes \texttt{MaxPooling} with \texttt{sum} reduction. Best results from each experimental group are marked in \textbf{bold} face.}
\label{tab}
\begin{tabular*}{0.87\textwidth}{l@{\extracolsep\fill}lll}
\hline
Loss              & Network &   Batch Size & R-AED (\(\uparrow\))  \\ \hline
Mean squared error (baseline)       & \texttt{Ave/mean}      & 8  & \textbf{5.69}  \\
Softmax cross entropy       & \texttt{Ave/mean}      & 8  & 3.45  \\
Multiscale softmax cross entropy    & \texttt{Ave/mean}      & 8  &  4.36  \\  \hline
Mean squared error (baseline)               & \texttt{Max/sum}     & 16    & 5.18 \\
Softmax cross entropy             & \texttt{Max/sum}     & 16  & 4.16 \\ 
Multiscale softmax cross entropy  & \texttt{Max/sum}     & 16  & \textbf{5.31} \\\hline
Mean squared error (baseline)    & \texttt{Max/sum}   & 8    & 5.53 \\
Softmax cross entropy             & \texttt{Max/sum}     & 8  & 4.99 \\ 
Multiscale softmax cross entropy  & \texttt{Max/sum}      & 8  & \textbf{6.12} \\\hline
\end{tabular*}
\end{table}

\section{Discussion}
Although the proposed MSCE loss has surpassed the commonly used MSE loss and the vanilla SCE loss based on the ablation experiments, some unstable predictions haven been noticed during the experiments. We assume that finetuned hyperparameters, including the weights \(\lambda_m\) in Equation~\ref{msce}, could mitigate this issue. 

In practice, the fovea is usually localized with the help of the relative position of the optic disc by surgeons. Therefore, it is expected that fusing the relative spatial information via optic disc segmentation would achieve better results for fovea localization. Additionally, the segmentation-based feature map of our approach could further strengthen the advantages by combining different ophthalmic tasks with fovea localization based on fundus images, such as vessel segmentation, optic disc and optic cup segmentation, and disease (\textit{e.g.} glaucoma) grading.

\section{Conclusion}
This work addresses the fovea localization task based on the feature map that is initially tailored for segmentation. Furthermore, the task of coordinate regression from logits is performed based on a probabilistic loss, which usually contributes to classification tasks. The modified multiscale version of softmax cross entropy (MSCE) has empirically shown the capability for localization tasks. The performance of MSCE surpasses both the vanilla SCE and the mean squared error loss with the identical network backbone and hyperparameter setups, which offers a novel loss alternative for fovea localization and is promising for other general coordinate regression tasks like bounding boxes in object detection. 

\subsubsection*{Acknowledgements.} This work was funded by the Deutsche Forschungsgemeinschaft (DFG, German Research Foundation) – grant 424556709/GRK2610.

\bibliographystyle{bvm}

\bibliography{3208}

\marginpar{\color{white}E\articlenumber}

\end{document}